\input{aipcheck}

\documentclass[
    ,final            
  ]
  {aipproc}
\usepackage{amsmath}
\layoutstyle{8x11single}

\newcommand{\ima}{{\mbox{Im}\,}}

\begin{document}

\title{$m_\pi$ and $N_c$ dependence of resonances form Unitarized Chiral Perturbation Theory}


\author{G. R\'ios}{
  address={Departamento de F\'isica Te\'orica II. Universidad Complutense de Madrid.}
}

\author{C. Hanhart}{
  address={Institut f\"ur Kernphysik and J\"ulich Center for Hadron 
    Physics, Forschungzentrum J\"ulich GmbH.}
}

\author{J. R. Pel\'aez}{
  address={Departamento de F\'isica Te\'orica II. Universidad Complutense de Madrid.}
}

\begin{abstract}
We review our studies on the $\rho$ and $\sigma$ resonances properties derived from
the Inverse Amplitude Method.
In particular, we study the leading $1/N_c$ behavior of the
resonances masses and widths and their evolution with changing $m_\pi$.
The $1/N_c$ expansion gives a clear definition of $\bar qq$ states, which is
neatly satisfied by the $\rho$ but not by the $\sigma$, showing that its
\emph{dominant component} is not $\bar qq$. The $m_\pi$ dependence of the resonance
properties is relevant to connect with lattice studies. We show that our
predictions compare well with some lattice results and we find that the 
$\rho\pi\pi$ coupling constant is $m_\pi$ independent, in contrast with
the $\sigma\pi\pi$ coupling, that shows a strong $m_\pi$ dependence.    
\end{abstract}

\maketitle

Light hadron spectroscopy lies beyond the realm of perturbative QCD. At low
energies, however, one can use the QCD low energy effective theory, named Chiral
Perturbation Theory (ChPT) \cite{chpt1}, 
to describe the dynamics of the lightest mesons. ChPT
describes the interactions of the pseudo Nambu-Goldstone bosons of the QCD 
chiral symmetry breaking, namely, the pions, by means of a 
effective lagrangian compatible with all QCD symmetries involving
only the pion field. The infinite tower of terms in this lagrangian is
organized as a low energy expansion in powers of $p^2/\Lambda_\chi^2$,
where $p$ stands either for derivatives, momenta or masses, and 
$\Lambda_\chi\simeq 4\pi f_\pi$, where $f_\pi$ denotes the pion decay constant.
ChPT is renormalized order by order by absorbing loop divergences in the 
renormalization of higher order parameters, known as low energy constants (LECs),
that parametrize the high energy QCD dynamics and \emph{carry no energy or mass
dependence}. They depend on a regularization scale $\mu$ but after 
renormalization the observables are independent of this scale. The value
of the LECs depend on the underlying QCD dynamics and are determined from
experiment. Up to the desired order, the ChPT expansion provides a 
\emph{systematic and model independent} description of how observables depend
on some QCD parameters like the light quark mass $\hat m=(m_u+m_d/2)$ or the
number of colors, $N_c$ \cite{'tHooft:1973jz}. 

The use of ChPT is limited to low energies and masses, nevertheless, combined with
dispersion relations and elastic unitarity it leads to a successful description of
meson dynamics up to energies around 1 GeV, generating resonant states
not originally present in the lagrangian,
without any a priori assumption on their existence or nature. In particular, we
find the $\rho$ and $\sigma$ resonances as poles on the second Riemann sheet of 
$\pi\pi$ elastic scattering amplitudes. 
With this approach we can then 
study some of these resonances properties, 
like their spectroscopic nature through their mass
and width dependence on $N_c$, or their dependence on the pion mass in order to 
connect with lattice studies. In the following sections we review this
``unitarized ChPT'' approach, named the Inverse Amplitude Method (IAM)~
\cite{Truong:1988zp,Dobado:1996ps,Guerrero:1998ei}, 
and then apply it to study the leading $1/N_C$ behavior and the chiral extrapolation
of the $\rho$ and $\sigma$ mesons.

The $\rho$ and $\sigma$ resonances appear as poles on the second Riemann sheet of
the $(I,J)=(1,1)$ and $(I,J)=(0,0)$ $\pi\pi$ scattering partial waves of definite
isospin, $I$ and angular momentum $J$, respectively. Elastic unitarity implies for
these partial waves, $t(s)$, and physical values of $s$ below inelastic thresholds,
that
\begin{equation}
  \label{unit}
  \ima t(s)=\sigma (s) \vert t(s)\vert^2 
  \;\;\Rightarrow\;\; 
  \ima\frac1{t(s)}=-\sigma(s),\qquad {\rm with}
  \quad \sigma(s)=2p/\sqrt{s},
\end{equation}
where $s$ is the Mandelstam variable and $p$ is the center of mass momentum.
Consequently, the imaginary part
of the inverse amplitude is known exactly. However, 
ChPT amplitudes, being an expansion
$t\simeq t_2+t_4+\cdots$, with
$t_k=O(p^k)$, can only satisfy
Eq. (\ref{unit}) perturbatively
\begin{equation}
  \label{unitpertu}
  \ima t_2(s)=0,\;\;\;
  \ima t_4(s)=\sigma(s) t_2^2(s)\;\;\dots
\end{equation}
and cannot generate poles. Therefore the resonance region lies
beyond the reach of standard ChPT. This region however, can be
reached combining ChPT with dispersion theory through the 
IAM~\cite{Truong:1988zp,Dobado:1996ps,Guerrero:1998ei}. 

The analytic structure of the $\pi\pi$ scattering amplitude $t(s)$,
consisting on a right cut extending from $s_{th}=4m_\pi^2$ to $\infty$,
and a left cut from $-\infty$ to 0,
allows to write a dispersion relation for the auxiliary function
 $G(s)\equiv t_2^2(s)/t(s)$
\begin{equation}
  \label{1/tdisp}
  G(s)=G(0)+G'(0)s+\tfrac{1}{2}G''(0)s^2+\frac{s^3}{\pi}
  \int_{s_{th}}^{\infty}\,ds'\frac{\ima G(s')}{s'^3(s'-s-i\epsilon)}+
  LC(G)+PC,
\end{equation}
where the integral over the left cut has been abbreviated as $LC(G)$ and $PC$
stands from possible pole contributions corresponding zeros of $t$.
The different terms in Eq.\eqref{1/tdisp} can be evaluated using
unitarity and ChPT as follows:
The right cut can be \emph{exactly} evaluated
  taking into account the elastic unitarity conditions
  Eqs.~(\ref{unit}),~(\ref{unitpertu}),
  $\ima G(s')=-\sigma(s')t_2^2(s')=-\ima t_4(s')$,
  for $s'\in (4m_\pi^2,\infty)$.
The subtraction constants only involve the
  amplitude and its derivatives evaluated at $s=0$,
  so they can be safely approximated with ChPT:
  $G(0)\simeq t_2(0)-t_4(0)$, $G'(0)\simeq t'_2(0)-t'_4(0)$,
  $G''(0)\simeq -t''_4(0)$.
The left cut, which is suppresed by $1/s'^3(s'-s)$, is 
  \emph{weighted at low energies},
  so it is appropriate to approximate it with ChPT: 
  $LC(G)\simeq -LC(t_4)$.
The pole contribution only appears in the scalar wave,
  which vanishes at the so called Adler zero. It counts $O(p^6)$ and
  and has been calculated explicitly \cite{modIAM}
  and is not just formally suppressed, but numerically negligible
  except near the Adler zero, away from the physical region.

Neglecting $PC$ for the moment, and taking into account that $t_2(s)$ is
just a first order polynomial in $s$, and that a dispersion
relation can be also written for $t_4$, we can write Eq.\eqref{1/tdisp}
as
\begin{equation}
  \label{1/tdispeval}
  G(s)\equiv\frac{t_2^2(s)}{t(s)}\simeq t_2(0)-t'_2(0)s
    -t_4(0)-t'_4(0)s-\tfrac1{2}t''_4(0)s^2-
    \frac{s^3}{\pi}
    \int_{s_{th}}^{\infty}\,ds'\frac{\ima\, t_4(s')}{s'^3(s'-s-i\epsilon)}-LC(t_4)
    =t_2(s)-t_4(s),
\end{equation}
which immediately leads to the IAM formula
$t^{IAM}(s)=\frac{t_2^2(s)}{t_2(s)-t_4(s)}.$
The IAM formula satisfies exact elastic unitarity and,
when reexpanded at low energies, reproduces the
ChPT expansion up to the order used to approximate the subtraction
constants and the left cut. Here we have presented an
$O(p^4)$ IAM but it can be generalized to higher chiral orders.
 Note that in the IAM derivation ChPT has
been always used at low energies, to evaluate parts of a
dispersion relation whose elastic unitarity cut has been taken into 
account exactly. Thus, there are no additional model dependencies in the approach,
which is reliable up to energies where
inelasticities become important. Taking the 
pole contribution into account leads to a modified IAM formula
\cite{modIAM} which is
almost indistinguishable from the ordinary one except in the
Adler zero region, where it fixes some problems of the 
ordinary IAM with the Adler zero \cite{modIAM}. Actually, this modified IAM
formula is the one used in this work since, as it will be shown below,
one amplitude pole gets near the Adler zero region.

This simple IAM formula is able to reproduce $\pi\pi$ scattering phase shift data
up to roughly 1 GeV and generates the poles associated
to the $\rho$ and $\sigma$ resonances with values of the LECs compatible 
with standard ChPT \cite{Guerrero:1998ei}. The $1/N_c$ expansion 
is implemented in ChPT through the LECs, whose leading $1/N_c$ scaling is known from QCD.
Also, the quark mass dependence implemented in the IAM agrees with that of ChPT up
to the order used. Hence, it is straightforward to study the leading $1/N_C$
behavior and the $\hat m$ dependence of the resonances generated with the IAM,
which we proceed to expose in the following sections. 

\section{Nature of resonances from their leading $1/N_c$ behavior}
The QCD $1/N_c$ expansion \cite{'tHooft:1973jz} provides a clear definition of
$\bar qq$ bound states: their masses and widths scale as $O(1)$ and $O(1/N_c)$
respectively. The QCD leading $1/N_c$ behavior of the ChPT parameters
($f_\pi$, $m_\pi$ and the LECs) is well known. Hence, by scaling with 
$N_c$ the ChPT parameters in the IAM, the $N_c$ dependence of the
$\rho$ and $\sigma$ mesons mass and width has been determined 
\cite{Pelaez:2003dy,Pelaez:2006nj}. They are defined from the pole positions as 
$\sqrt{s_{pole}}=M-i\Gamma$. Note that we should not take too large
$N_C$ values, since the $N_c\to\infty$ is a weakly interacting limit,
where the IAM approach is less reliable \cite{Pelaez:2009eu}. 
Also, for very large $N_c$ values,
even a tiny admixture of $\bar qq$ in the physical state would become dominant,
but this does not give any information about the dominant component of the 
$N_c=3$ physical state.

\begin{figure}[t]
  \centering
  \hbox{
    \includegraphics[angle=-90,scale=.35]{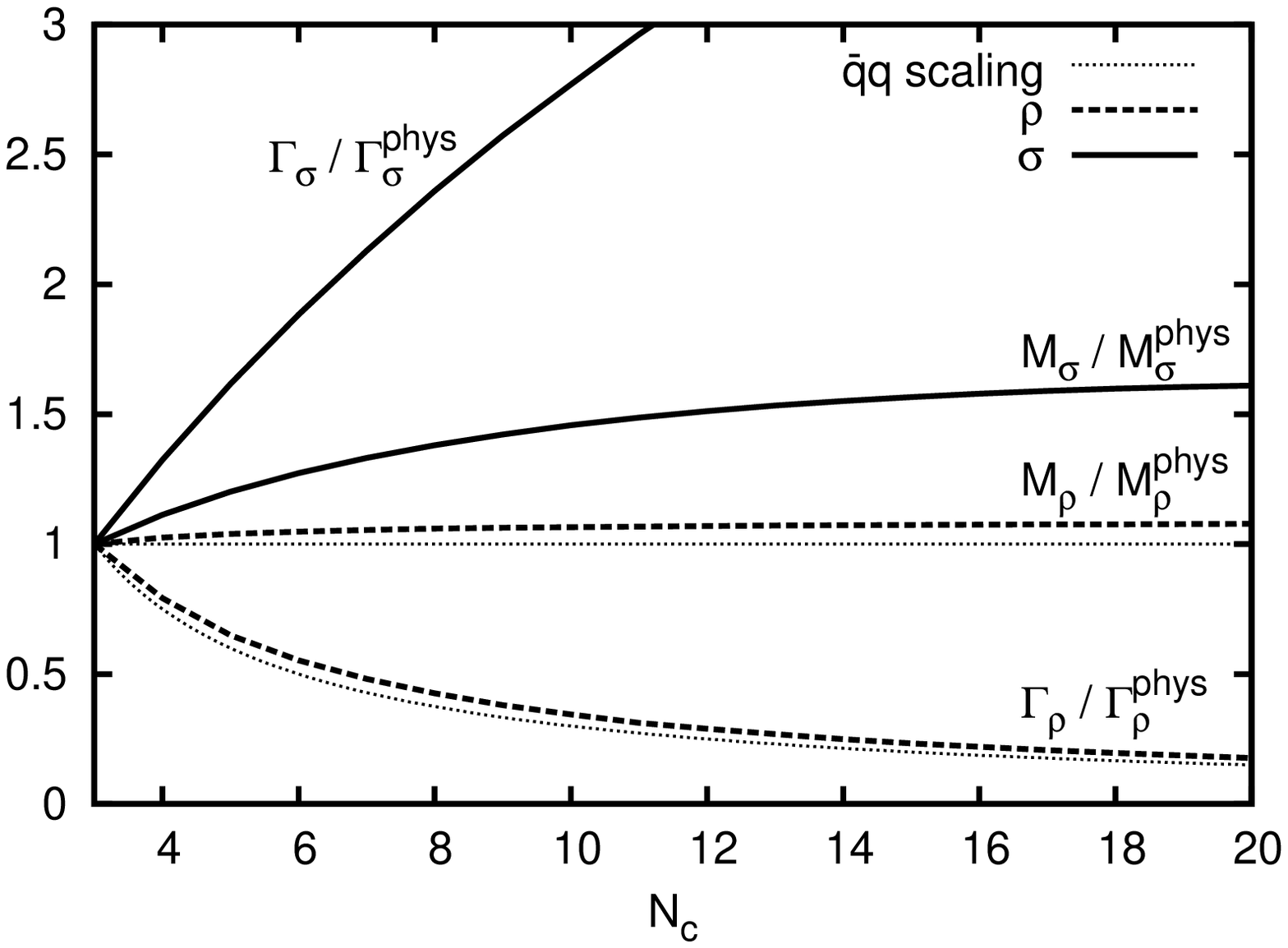}
    \includegraphics[angle=-90,scale=.35]{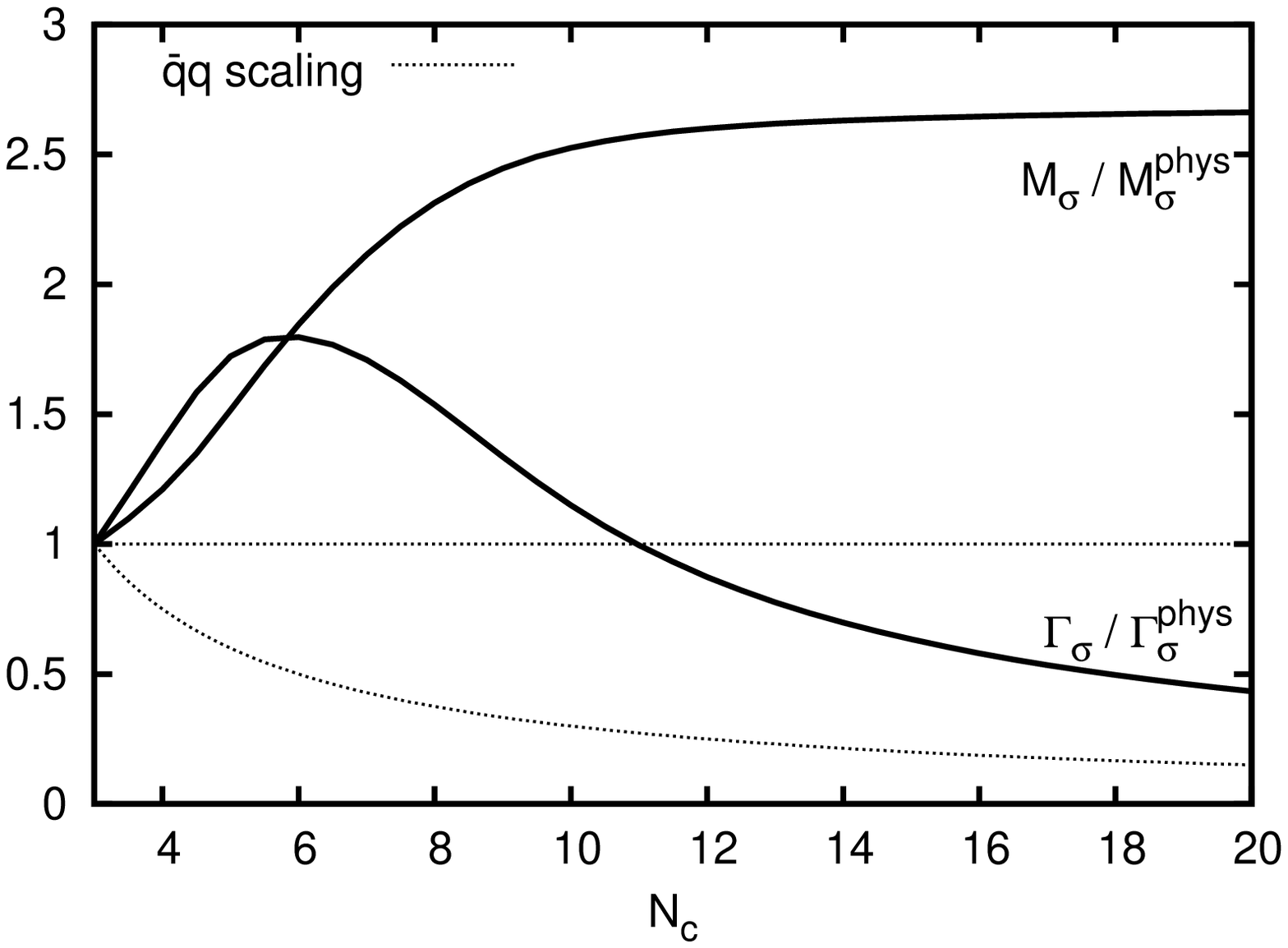}
  }
  \caption{{\bf Left:} $\rho$ and $\sigma$ $1/N_c$ scaling $O(p^4)$. 
    {\bf Right:} $\sigma$ $1/N_c$ scaling $O(p^6)$.}
  \label{ncscaling}
\end{figure}

Fig. \ref{ncscaling} (left) shows the $\rho$ and $\sigma$
mass and width $N_c$ scaling. It can be clearly seen that 
the $\rho$ follows remarkably well the
expected behavior of a $\bar qq$ state, confirming that the method
can obtain the correct $N_c$ behavior of well known $\bar qq$ states.
In contrast,the $\sigma$ does not that $\bar qq$ pattern, 
allowing us to conclude that \emph{the $\sigma$ dominant component is not
$\bar qq$}. 

Loop contributions play an important role in determining the
$\sigma$ pole position. Since they are $1/N_c$ suppressed compared to 
tree level terms, it may happen that for larger $N_c$ values they
become comparable to tree level $O(p^6)$ contributions,
which are subdominant in the chiral expansion, but not $N_c$
suppressed. Thus we checked the $O(p^4)$ results with an
$O(p^6)$ IAM calculation \cite{Pelaez:2006nj}. We defined a 
$\chi^2$-like function to measure how close a resonance is 
from a $\bar qq$ behavior. First, we used it at $O(p^4)$ to show
that it is not possible for the $\sigma$ to behave predominantly as 
a $\bar qq$ state describing simultaneously the data and the $\rho$
$\bar qq$ behavior. Next, we obtained an $O(p^6)$ data fit where the
$\rho$ $\bar qq$ behavior was imposed. Figure \ref{ncscaling} 
(right) shows
the $M_\sigma$ and $\Gamma_\sigma$ $N_c$ scaling obtained from that fit. 
Note that both $M_\sigma$ and $\Gamma_\sigma$ grow \emph{near $N_c=3$},
confirming the $O(p^4)$ result of a non $\bar qq$ dominant component.
However, for $N_c$ between 8 and 15, where we still trust the IAM,
$M_\sigma$ becomes constant and $\Gamma_\sigma$ starts decreasing. This may
hint to a \emph{subdominant $\bar qq$ component}, arising as loops become
suppressed as $N_c$ grows. Finally, by forcing the $\sigma$ to behave as a 
$\bar qq$, we found that in the best case this subdominant component could
become dominant around $N_c>6-8$, but always with an $N_c\to\infty$ mass
above 1 GeV instead of its physical $\sim$ 450 MeV value. This supports
the emerging picture of two low energy scalar nonets, 
one of exotic nature below 1 GeV and another of ordinary $\bar qq$
nature above 1 GeV.

\section{Chiral Extrapolation of the $\rho$ and $\sigma$ resonances}
ChPT also provides an expansion of $m_\pi$ in terms of $\hat m$
(at leading order $m_\pi^2\sim \hat m$). Thus, by changing $m_\pi$ in the
amplitudes we see how the IAM poles depend on $\hat m$.
We report here our analysis of the $\rho$ and $\sigma$ properties
dependence on $m_\pi$~\cite{Hanhart:2008mx}.

The values of $m_\pi$ considered should fall within the ChPT
applicability range and allow for some elastic 
regime below $K\bar K$, that would almost disappear if
$m_\pi>500$, which would be the most optimistic applicability range. 
We expect higher order
corrections to be more
relevant as $m_\pi$ increases. Thus, our results become less
reliable as $m_\pi$ grows.
 
\begin{figure}[t]
   \centering
   \vbox{
     \hbox{
       \includegraphics[scale=0.25,angle=0]{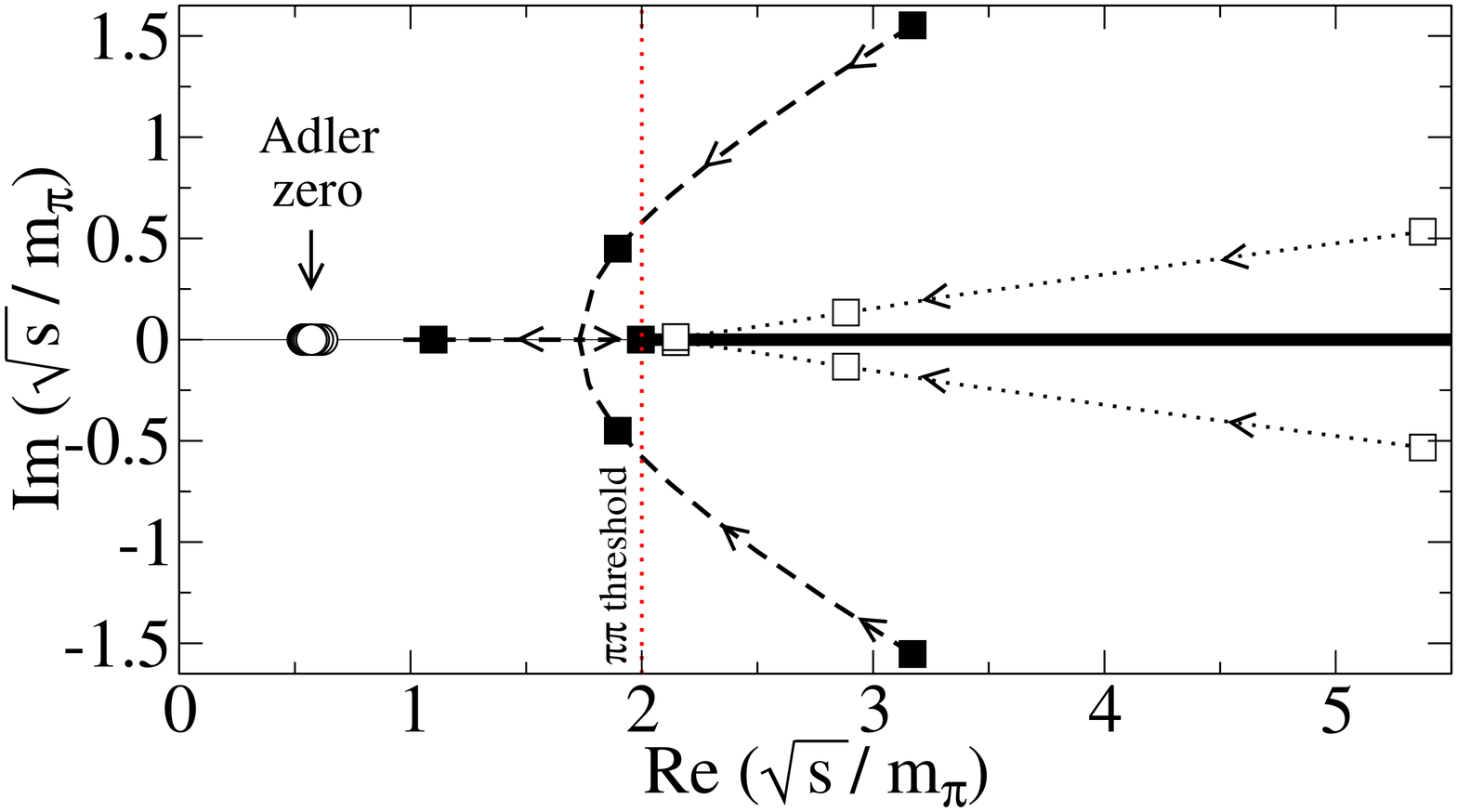}
       \includegraphics[scale=0.57,angle=0]{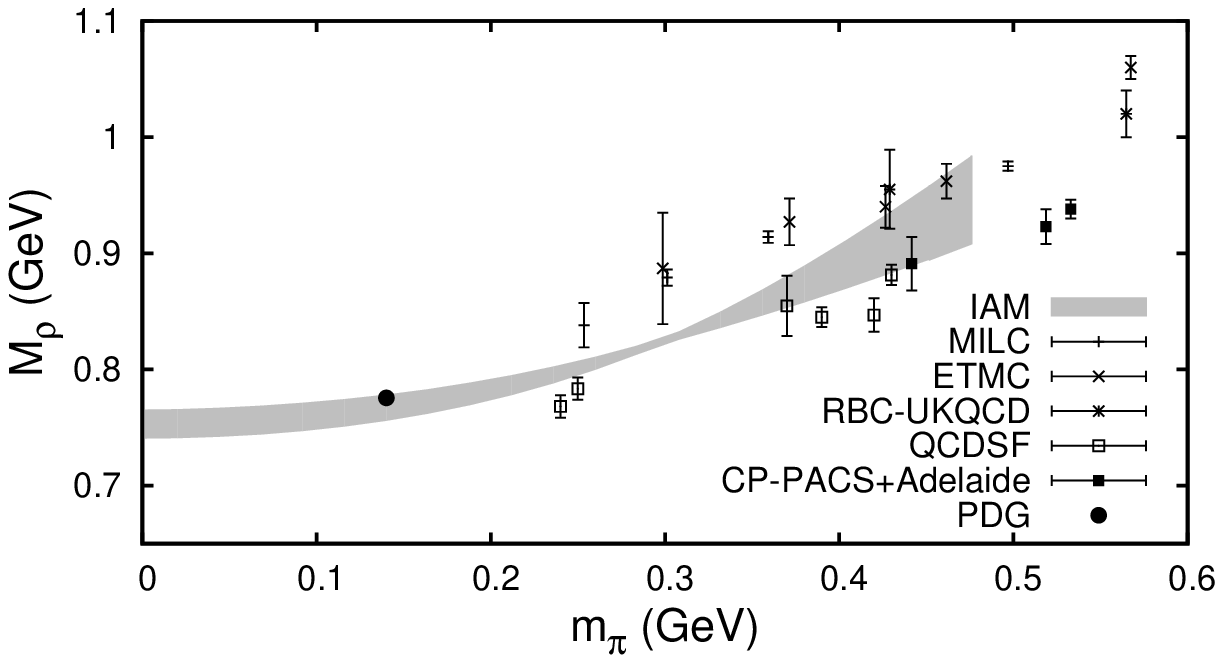}
     }
     \hbox{
       \includegraphics[scale=.35,angle=-90]{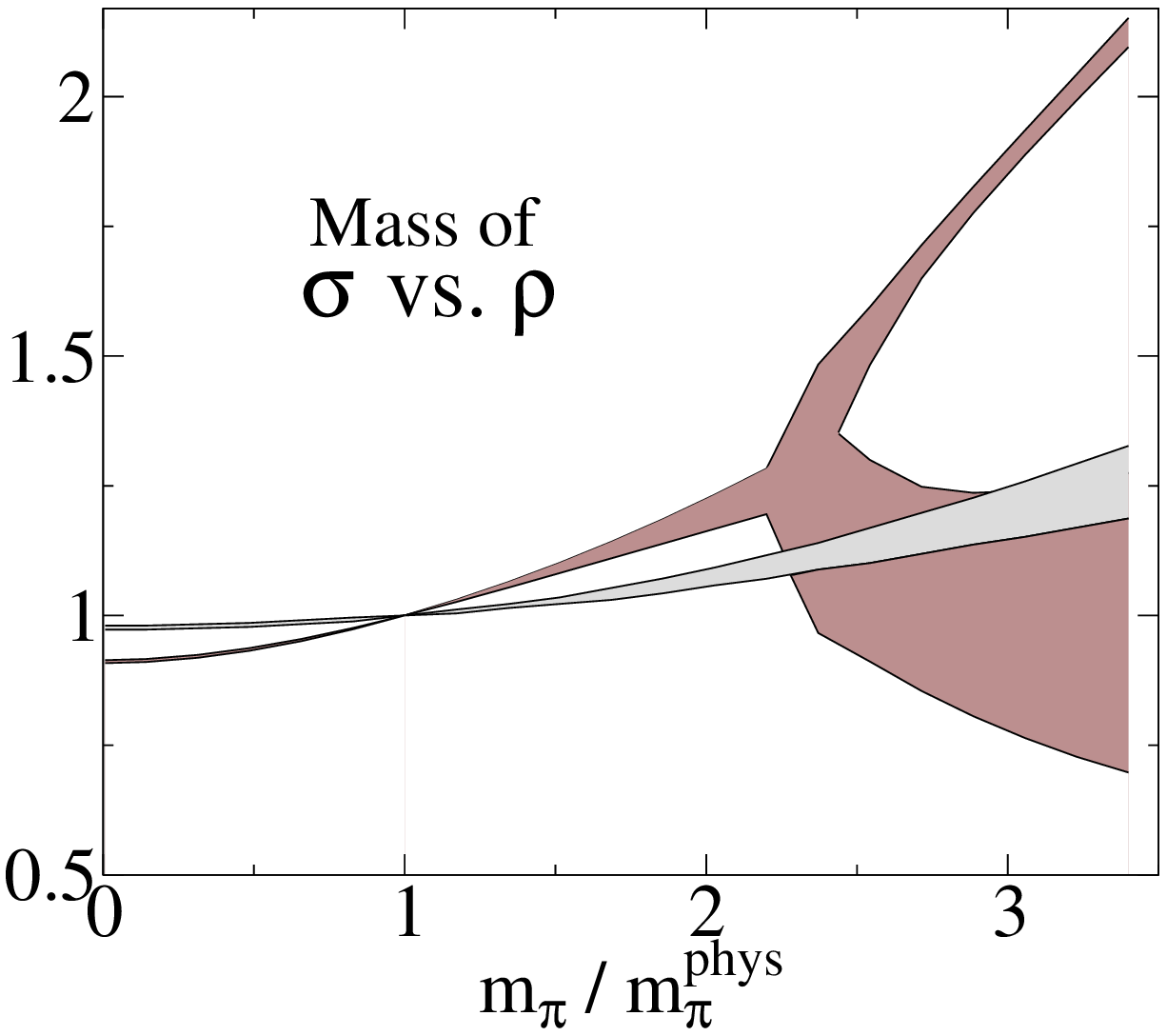}
       \includegraphics[scale=.35,angle=-90]{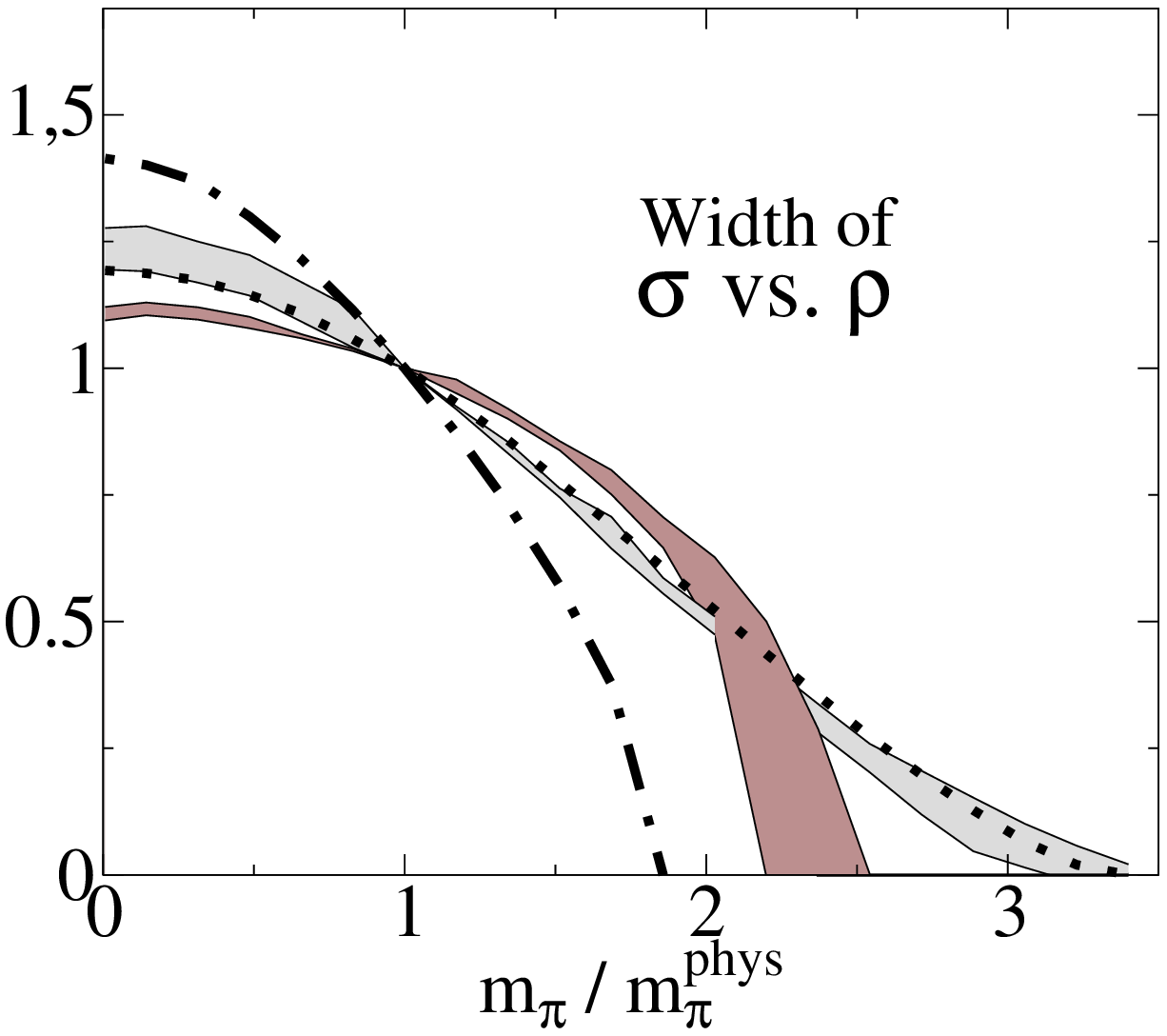}
       \includegraphics[scale=.30,angle=-90]{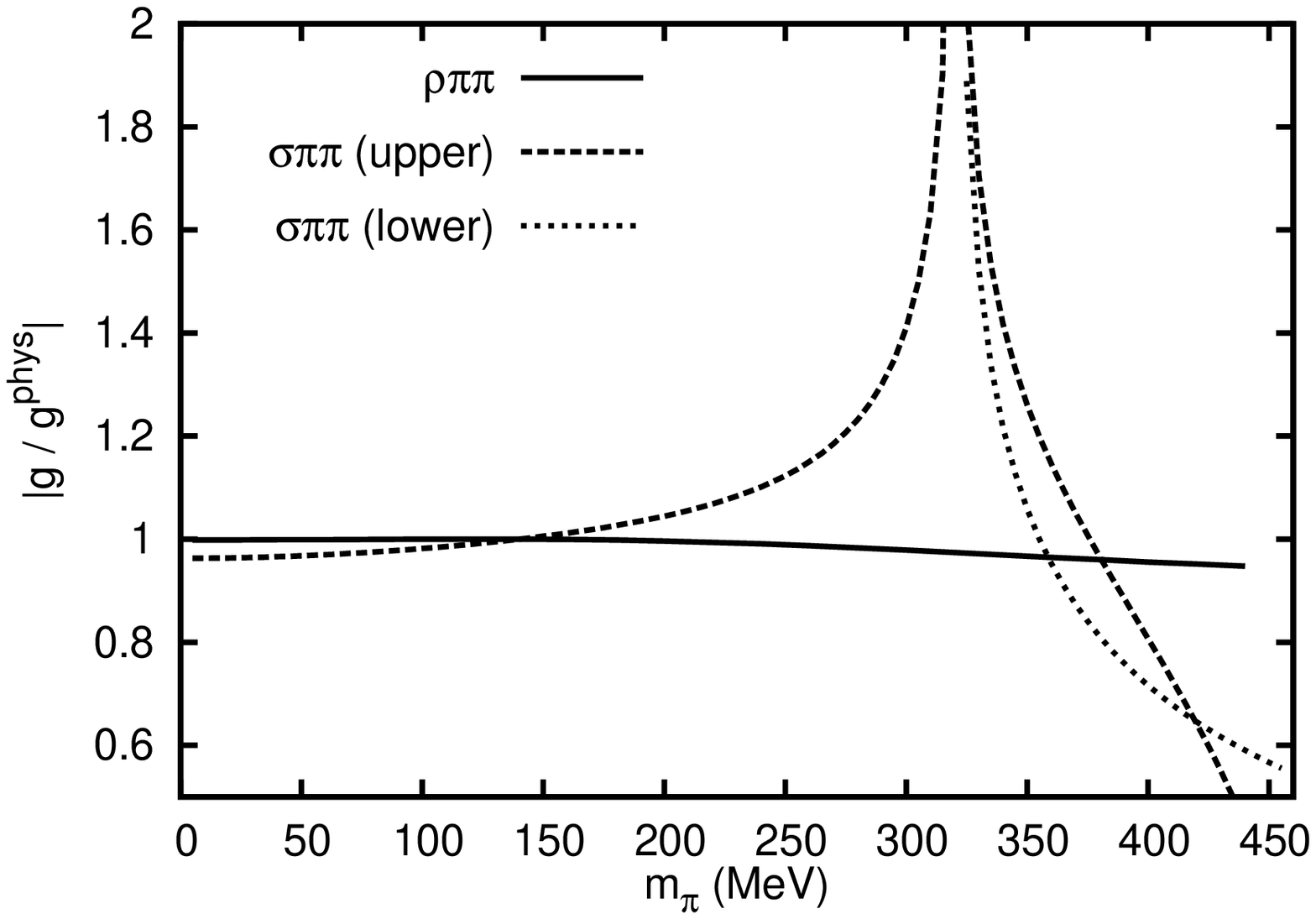}
     }
   }
     \caption{ {\bf Top Left:} Movement of the $\sigma$ (dashed lines) 
       and $\rho$ (dotted
       lines) poles for increasing $m_\pi$ (direction indicated by the
       arrows) on the second sheet.  The filled (open) boxes denote the
       pole positions for the $\sigma$ ($\rho$) at pion masses $m_\pi=1,\
       2,$ and $3 \times m_\pi^{\rm phys}$, respectively. 
       {\bf Top Right:} Comparison the IAM $M_\rho$
       dependence on $m_\pi$ with some recent lattice results\cite{lattice1}. 
       {\bf Bottom Left:} Comparison of the $\rho$ (light) and $\sigma$
       (dark) mass dependence on $m_\pi$. 
       {\bf Bottom Center:} Comparison of the $\rho$ (light) and $\sigma$
       (dark) width dependence on $m_\pi$.
       The dotted ($\rho$) and  dot-dashed ($\sigma$) lines show the 
       decrease due to only phase space
       assuming a constant coupling to $\pi\pi$.
       {\bf Bottom Right:} $\rho$ and $\sigma$ couplings calculated from
       the pole residue. In all panels, the bands cover the LECs uncertainty.}  
       \label{massandwidth}
 \end{figure}
Fig. \ref{massandwidth} (top left) shows the evolution of the $\sigma$ and $\rho$
pole positions as $m_\pi$ is increased. In order to see the pole
movements relative to the $\pi\pi$ threshold, which is also increasing,
we use units of $m_\pi$, so the threshold is
fixed at $\sqrt{s}=2$. Both poles move closer to threshold and
they approach the real axis. The $\rho$ poles reach the real axis
at the same time that they cross threshold.
One of them jumps into the first sheet and becomes a bound state, 
while its conjugate partner remains on the second sheet practically at the very same
position as that in the first. In contrast, the $\sigma$
poles go below threshold with a finite imaginary part before they
meet in the real axis, still on the second sheet, becoming
virtual states. As $m_\pi$ increases, one pole
moves toward threshold and jumps through the branch point to the
first sheet staying in the real axis below threshold, very
close to it as $m_\pi$ keeps growing. The other $\sigma$ pole moves
down in energies away from threshold and remains 
on the second sheet.
Similar movements were found within quark models
\cite{vanBeveren:2002gy} and finite temperature and 
density analysis 
\cite{Patkos:2002vr,FernandezFraile:2007fv}.

Fig. \ref{massandwidth} (top right) shows our results for the $\rho$ mass
dependence on $m_\pi$ compared with some lattice 
results~\cite{lattice1}, and the
PDG value for the $\rho$ mass. 
Now $M_\rho$ is defined as the point where the phase shift crosses $\pi/2$,
except for those $m_\pi$ values where the $\rho$ becomes a bound
state, where it is defined from the pole position.
In view of the incompatibilities between
different lattice collaborations, we find a qualitative good
agreement with lattice results. 
The $M_\rho$ dependence on $m_\pi$
agrees also with estimations 
for the two first coefficients of its chiral expansion \cite{bruns}.

In Fig. \ref{massandwidth} (bottom left) we compare the $m_\pi$ dependence
of $M_\rho$ and $M_\sigma$, normalized to their physical values.
The bands cover the LECs uncertainties. Both masses
grow with $m_\pi$, but $M_\sigma$ grows faster than $M_\rho$. 
Above $2.4 \, m_\pi^{\rm phys} $, we show two bands since the
two $\sigma$ poles lie on the real axis with two different masses. 

In the bottom center panel of Fig. \ref{massandwidth} we compare the $m_\pi$
dependence of $\Gamma_\rho$ and $\Gamma_\sigma$ normalized to their
physical values: note that both widths become smaller. 
We compare this decrease with the expected phase space reduction 
 as resonances approach the $\pi\pi$ threshold.
We find that $\Gamma_\rho$ follows very well
this expected behavior, which
implies that the $\rho\pi\pi$ coupling is almost $m_\pi$ independent.
In contrast, $\Gamma_\sigma$ deviates from the
phase space reduction expectation. This suggests a strong $m_\pi$
dependence of the $\sigma$ coupling to two pions, which we
confirm with a explicit calculation of the resonances couplings from
the pole residues as shown in the bottom left panel.



\bibliographystyle{aipproc}   

\bibliography{sample}

\IfFileExists{\jobname.bbl}{}
 {\typeout{}
  \typeout{******************************************}
  \typeout{** Please run "bibtex \jobname" to optain}
  \typeout{** the bibliography and then re-run LaTeX}
  \typeout{** twice to fix the references!}
  \typeout{******************************************}
  \typeout{}
 }


\end{document}